\documentclass{aa}
\usepackage{txfonts}
\usepackage{natbib}
\usepackage{tikz}
\usetikzlibrary{decorations.pathmorphing}
\usetikzlibrary{decorations.pathmorphing,calc}
\def\ut#1{\mathop{\vtop{\ialign{##\crcr
     $\hfil\displaystyle{#1}\hfil$\crcr\noalign
     {\kern1pt\nointerlineskip}\hbox{$\hfil\sim\hfil$}\crcr
     \noalign{\kern1pt}}}}}

\def\undersymbol#1#2{\mathop{\vtop{\ialign{##\crcr
     $\hfil\displaystyle{#2}\hfil$\crcr\noalign
     {\kern1pt\nointerlineskip}\hbox{$\hfil#1\hfil$}\crcr
     \noalign{\kern1pt}}}}}

\usepackage{graphicx}

\begin{document}

\title{Hubble tension and absolute constraints on the local Hubble parameter}
       \author{V.G. Gurzadyan\inst{1,2},
       A. Stepanian\inst{1}}
       
              \institute{Center for Cosmology and Astrophysics, Alikhanian National Laboratory and Yerevan State University, Yerevan, Armenia \and
                               SIA, Sapienza University of Rome, Rome, Italy}

   \offprints{V.G. Gurzadyan, \email{gurzadyan@yerphi.am}}
   \date{Submitted: XXX; Accepted: XXX}

 \abstract{
It is shown, from the two independent approaches of McCrea-Milne and of Zeldovich, that one can fully recover the set equations corresponding to the relativistic equations of the expanding universe of Friedmann-Lemaitre-Robertson-Walker geometry. Although similar, the Newtonian and relativistic set of equations have a principal difference in the content and hence define two flows, local and global ones, thus naturally exposing the Hubble tension at the presence of the cosmological constant $\Lambda$. From this, we obtain  absolute constraints on the lower and upper values for the local Hubble parameter, $\sqrt{\Lambda c^2/3} \simeq 56.2$ and $\sqrt{\Lambda c^2} \simeq 97.3$ (km/sec Mpc$^{-1}$), respectively. The link to the so-called maximum force--tension issue in cosmological models is revealed.}


   \authorrunning{V.G. Gurzadyan, A. Stepanian}
   \titlerunning{Hubble tension and absolute constraints on the local Hubble parameter}
   \maketitle
%

\section{Introduction}

Non-relativistic cosmological schemes,  developed a long time ago \citep{MM1,MM2}, are
considered certainly efficient, such as the Newtonian limit of General Relativity, which has a  limited scale for application. 
Moreover, according to  \citet{Z}, the Newtonian approach would have enabled us to predict the expansion of the Universe far earlier, and he considered it ``paradoxical that the Newtonian theory of  expansion was discovered only after the achievement of Friedmann''.

Non-relativistic cosmology as mentioned in textbooks (e.g., \citealt{MTW}), had no  link to the cosmological constant, which had been typically considered negligible or zero for many decades.   
In \cite{G} it was shown that the cosmological constant $\Lambda$ can be naturally introduced in the non-relativistic cosmology (i.e., in the McCrea-Milne scheme). We note that \cite{G} appeared before the discovery of the accelerated expansion of the Universe, and hence the emergence of the role of the cosmological constant. The generalized function satisfying Newton's basic principle of the equivalency of the gravity of a  sphere and point mass  was derived in \cite{G}, which contains one more term, exactly of the  cosmological constant form (see discussion in \citealt{Ch}). The emerged metric for the weak-field General Relativity (GR) at spherical symmetry is \citep{GS1,G1}

\begin{equation}\label{mod}
g_{00} = 1- \frac{2GM}{c^2 r} - \frac{\Lambda r^2}{3}; \quad g_{rr} = (1- \frac{2GM}{c^2 r} -\frac{\Lambda r^2}{3})^{-1}\,.
\end{equation}

This metric is of Schwarzschild--de Sitter form, and as a weak-field limit of GR can be applied to the dynamics of groups and clusters of galaxies \citep{GS1,G1,GS2}.
In Eq.(1) the cosmological constant $\Lambda$ acts as a second fundamental constant \citep{GS1} to describe gravity, along with Newton's gravitational constant $G$. 

Consideration of $\Lambda$ as the fourth fundamental constant,  along with $c$, $\hbar$, and $G$, enables us to construct a class of dimensionless quantities as  \citep{GS4} 
\begin{equation}\label{Nat}
m I^n= m \frac{c^{3n}}{\hbar^n G^n \Lambda^n}, \quad m, n \in \mathbb{R}\,,
\end{equation}
which at certain values of $n$ and $m$ leads to an information scaling \citep{GS4,GSbh} and is linked to the  Bekenstein bound \citep{BB}. 

This approach,  the weak-field modified GR with Eq.(1), provides an explanation to the Hubble tension \citep{VTR,R,Val},  the discrepancy between the {local} and {global} values of the Hubble parameter, without any free parameter \citep{GSt1,GSt2}. Among the approaches dealing with the Hubble tension (see the extensive references in \citealt{Val}), we mention that of \cite{photon}, again without a free parameter, which deals with the basic physical principles. Clearly, the further revealing of the systematics in the different observational techniques is required when clarifying the Hubble tension issue and its possible link to  basic physics.

In our approach we deal with two different flows: {global} and {local}. The global flow is related to the Friedmann-Lemaitre-Robertson-Walker (FLRW) geometry, while the local flow occurring at the non-relativistic regime is a result of the repulsive nature of the $\Lambda$ term in the equations of gravity according to Eq.(\ref{mod}). Accordingly, the data of the \textit{Planck} satellite \cite{Pl2018}  refer to the measurements of the {global} Hubble parameter, while those obtained  via the {local} distance indicators  (e.g., \citealt{RApJ}) are assigned to the {local} flow as 
\begin{equation}\label{Hflowsvalues}
Planck : H_{global}=67.4 \pm 0.5 \quad  kms^{-1}Mpc^{-1}
,\end{equation}
\begin{equation}
HST : H_{local}= 74.03 \pm 1.42 \quad kms^{-1}Mpc^{-1}.
\end{equation}

It should be noted, on the one hand, that the modified gravity is among the approaches used to describe   dark matter and large-scale matter distribution; on the other hand,  observations are efficiently used to constrain the deviations from GR (see, e.g., \citealt{KEK,GK,E,Cap3,Ch}).  

In this paper we show that {energy} (McCrea-Milne) and {force} (Zeldovich) approaches together enable us to define the complete set of non-relativistic equations corresponding to the Friedmann equations. Then, each set of equations defines a flow,  however with different content and Hubble parameters. We show that the local Hubble flow defined by Eq.(1) leads to principal constraints on the upper and lower values of the Hubble parameter defining that flow. It is remarkable that these constraints are determined by $\Lambda$ as of the second constant of gravity entering Eq.(1). The same approach enables us to address the maximum force--tension issue in cosmology and to show the  consistency of  relativistic and non-relativistic definitions.

\section{Non-relativistic cosmology}

The non-relativistic cosmology as given in textbooks \citep{MTW}, also known as the McCrea-Milne cosmology \citep{MM1,MM2}, is based on the conservation of energy of a test particle on an expanding sphere with density $\rho$ and radius $r(t)$ . In the context of $\Lambda$-gravity, the evolution of a non-relativistic universe in this cosmological model is governed by  
\begin{equation}\label{MM1}
\frac{v^2(t)}{r^2(t)}= H^2 = \frac{8 \pi  G \rho}{3} + \frac{\Lambda c^2}{3}\,.
\end{equation}
Thus, this approach is based on the  {energy} relation, which coincides with the first Friedmann equation of the FLRW metric in relativistic cosmology. 

On the other hand, in his less known note \citet{Z} proposed his own approach for non-relativistic cosmology which is based on the notion of gravitational {force}. Similarly, in the context of $\Lambda$-gravity, his approach yields 

\begin{equation}\label{Zel}
\frac{d H}{dt} + H^2 = -\frac{4 \pi G \rho}{3} + \frac{\Lambda c^2}{3}\,.
\end{equation}   
We can see that  this momentum equation provides the second  Friedmann equation of FLRW cosmology for pressureless dust $\nabla P = 0$.
  
Then, considering both approaches,  {energy} and {force}, we obtain the full set of equations of the non-relativistic cosmology, which are similar to the relativistic FLRW equations. However, and it is of principal importance, the nature of the {local} Hubble constant $H(t)$ obtained in Eqs.(\ref{MM1} and \ref{Zel}), is drastically different from the {global} $H(t)$ obtained from the relativistic FLRW equations. If so, we might predict the Hubble tension (i.e., the difference in the values of the local and global Hubble flows). In the next section we show that the local flow enables us to obtain strict theoretical upper and lower limits for the value of the local Hubble parameter.

\section{Constraints on the local Hubble flow}

In \cite{GSt1,GSt2} it is proposed that the local Hubble flow is not due to the expansion of the Universe, but to the presence of the repulsive $\Lambda$ term in the equations of weak-field limit of GR (cf. \citealt{Ch}). Considering a test body of mass $m$ in the vicinity of a central body of mass $M$, the gravitational force acting on the test body in the context of this $\Lambda$-gravity Eq.(1) is written as  
\begin{equation}\label{force}
F = - \frac{GMm}{r^2} + \frac{\Lambda c^2 r m}{3}\,.
\end{equation}  
Then a critical radius $r_{crit}$ can be introduced as  
\begin{equation}\label{crit}
r_{crit}^3 = \frac{3GM}{\Lambda c^2},
\end{equation}
which determines the scale of dominance of the $\Lambda$ term. When the distance between two bodies is larger than $r_{crit}$, the test body starts to move away from the central object, which causes the local Hubble flow. So, in order to observe the local flow, the system under consideration should have a radius larger than $r_{crit}$. Therefore, the absolute value for the upper limit of $H$ for any system will be obtained at its $r_{crit}$  as  
\begin{equation}\label{ulimit}
H^2_{max}= (\frac{v^2}{r_{crit}^2}) = \frac{8 \pi  G \rho}{3}|_{r_{crit}} + \frac{\Lambda c^2}{3} =\frac{2GM}{r_{crit}^3} + \frac{\Lambda c^2}{3}=\Lambda c^2.
\end{equation}

The local Hubble flow can be determined based on Eq.(2) as information (entropy) of the system undergone by the local Hubble flow due to presence of the central object mass, where  $M$ is given by the Bekenstein bound \citep{BB} as  \citep{GS2}
\begin{equation}\label{BB}
I_{BB} \leq \frac{2 \pi M R c}{\hbar ln2}.
\end{equation}

Here, similar to the case of black holes (BHs) in the presence of $\Lambda$ \citep{GSbh}, we have to ask the following question: {What is (are) the characteristic radius(radii) $R$ of the system?

Clearly, in this case $R$ should be the corresponding radius or distance where the outgoing flow is analyzed.

Analyzing the dynamics of objects in the context of $\Lambda$-gravity we   get two characteristic radii. The critical radius
\begin{equation}
r_{crit} = (\frac{3GM}{c^2 \Lambda})^{\frac{1}{3}}
\end{equation}
is the absolute minimum radius at which the flow can occur. 

Then the largest event horizon (the cosmological horizon) of the object 
is obtained by solving the  equation
\begin{equation}\label{Horiz}
1- \frac{2 GM}{c^2 r} -\frac{\Lambda r^2}{3} = 0,
\end{equation}
and can be regarded as the limit where the weak-field limit is not valid anymore, and at larger radii  should be  considered relativistic effects. Specifically, that radius is 
\begin{equation}\label{cosmic}
r_{cosmic} = \frac{2}{\sqrt{\Lambda}} cos (\frac{1}{3} cos^{-1} (\frac{3  GM \sqrt{ \Lambda}}{c^2} )- \frac{\pi}{3}),
\end{equation}
which is smaller than the horizon of de Sitter (dS) spacetime:  
\begin{equation}\label{dS}
r_{cosmic} < \sqrt{\frac{3}{\Lambda}}.
\end{equation}

Here the essential conceptual point is that the notion of horizon obtained in Eq.(\ref{cosmic}) is valid for all objects and is not restricted  to the BHs. On the other hand, for (non-extreme) Schwarzschild-de Sitter (SdS) BHs, in addition to  the $r_{cosmic}$, we have another solution for Eq.(\ref{Horiz}) which is defined as
\begin{equation}\label{SdSEH1}
r_{1} = \frac{2}{\sqrt{\Lambda}} cos (\frac{1}{3} cos^{-1} (\frac{3  GM \sqrt{ \Lambda}}{c^2} )+ \frac{\pi}{3}),
\end{equation}
and is slightly larger than $\frac{2GM}{c^2}$. These two radii can be illustrated via the Penrose diagram, Fig.(\ref{Fig1}), where the zigzag lines denote the spacelike singularity regarding the SdS BH.
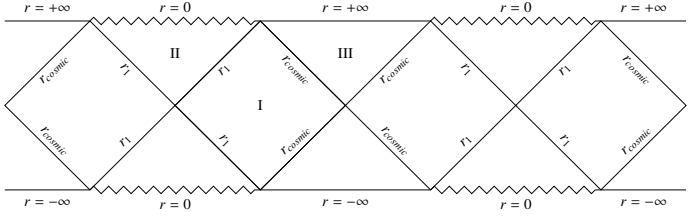
\begin{figure}
\scalebox{0.56}{
\centering 
\begin{tikzpicture} 
\node (I)    at ( 2,0)   {I};
\node (II)   at (-2,0)   {};
\node (III)  at (0, 1.25) {II};
\node (IV)   at (0,-1.25) {};
\node (V)   at (4,-1.25) {};
\node (VI)   at (4,1.25) {III};
\node (VII)   at (6,0) {};
\node (VIII)   at (8,1.25) {};
\node (IX)   at (8,-1.25) {};
\node(X) at (10,0) {};

\path  
  (II) +(90:2)  coordinate  (IItop)
       +(-90:2) coordinate (IIbot)
       +(0:2)   coordinate                  (IIright)
       +(180:2) coordinate(IIleft)
       ;
\draw (IIleft) -- 
          node[midway, below, sloped] {$r_{cosmic}$}
      (IItop) --
          node[midway, below, sloped] {$r_1$}
      (IIright) -- 
          node[midway, above, sloped] {$r_1$}
      (IIbot) --
          node[midway, above, sloped] {$r_{cosmic}$}
      (IIleft) -- cycle;

\path 
   (I) +(90:2)  coordinate (Itop)
       +(-90:2) coordinate  (Ibot)
       +(180:2) coordinate (Ileft)
       +(0:2)   coordinate (Iright)
       ;
\draw (Ileft) -- 
          node[midway, below, sloped] {$r_1$}
      (Itop) --
          node[midway, below, sloped] {$r_{cosmic}$}
      (Iright) -- 
          node[midway, above, sloped] {$r_{cosmic}$}
      (Ibot) --
          node[midway, above, sloped] {$r_1$}
      (Ileft) -- cycle;
\draw  (Ileft) -- (Itop) -- (Iright) -- (Ibot) -- (Ileft) -- cycle;

\path  
  (VII) +(90:2)  coordinate  (VIItop)
       +(-90:2) coordinate (VIIbot)
       +(180:2)   coordinate                  (VIIright)
       +(0:2) coordinate(VIIleft)
       ;
\draw (VIIleft) -- 
          node[midway, below, sloped] {$r_1$}
      (VIItop) --
          node[midway, below, sloped] {$r_{cosmic}$}
      (VIIright) -- 
          node[midway, above, sloped] {$r_{cosmic}$}
      (VIIbot) --
          node[midway, above, sloped] {$r_1$}
      (VIIleft) -- cycle;

\path  
  (X) +(90:2)  coordinate  (Xtop)
       +(-90:2) coordinate (Xbot)
       +(180:2)   coordinate                  (Xright)
       +(0:2) coordinate(Xleft)
       ;
\draw (Xleft) -- 
          node[midway, below, sloped] {$r_{cosmic}$}
      (Xtop) --
          node[midway, below, sloped] {$r_1$}
      (Xright) -- 
          node[midway, above, sloped] {$r_1$}
      (Xbot) --
          node[midway, above, sloped] {$r_{cosmic}$}
      (Xleft) -- cycle;

\draw[decorate,decoration=zigzag] (IItop) -- (Itop)
      node[midway, above, inner sep=2mm] {$r=0$};

\draw[decorate,decoration=zigzag] (IIbot) -- (Ibot)
      node[midway, below, inner sep=2mm] {$r=0$};

\draw (Ibot) -- (VIIbot)
      node[midway, below, inner sep=2mm] {$r=-\infty$};

\draw (Itop) -- (VIItop)
      node[midway, above, inner sep=2mm] {$r=+\infty$};

\draw[decorate,decoration=zigzag] (VIItop) -- (Xtop)
      node[midway, above, inner sep=2mm] {$r=0$};

\draw[decorate,decoration=zigzag] (VIIbot) -- (Xbot)
      node[midway, below, inner sep=2mm] {$r=0$};

\draw (Xtop) -- (12,2)
        node[midway, above, inner sep=2mm] {$r=+\infty$};
\draw (Xbot) -- (12,-2)
        node[midway, below, inner sep=2mm] {$r=-\infty$};
\draw (IItop) -- (-4,2)
        node[midway, above, inner sep=2mm] {$r=+\infty$};
\draw (IIbot) -- (-4,-2)
        node[midway, below, inner sep=2mm] {$r=-\infty$};

\end{tikzpicture}
}
\caption{Penrose diagram for SdS BH.} \label{Fig1}
\end{figure}

Consequently, for the local Hubble flow we  have
\begin{equation}\label{llimit}
H^2_{min}=  \frac{\Lambda c^2}{3}, 
\end{equation}
which defines the absolute lowest value for $H$ of the outgoing flow.

Thus, for any local flow we  have
\begin{equation}\label{limitflow}
I \leq \frac{2 \pi c}{\hbar ln 2} M r, \quad r_{crit}<r<r_{cosmic}.
\end{equation}

Now a natural question arises: {What will happen when these two limits coincide}:

$$r_{crit} = r_{cosmic}.$$

This condition occurs for an extreme SdS BH where the two (positive) event horizons $r_1$ and $r_2$ become identical,
\begin{equation}\label{hor}
r_{1} = r_{2} = \frac{1}{\sqrt{\Lambda}},
\end{equation}
and the mass of the BH is a combination of fundamental constants as
\begin{equation}\label{SdSm}
M_{SdS} = \frac{c^2}{3 G \sqrt{\Lambda}}\,.
\end{equation}
The following radius is  critical for such a BH \citep{SKG}: 
\begin{equation}\label{horcrit}
r_{crit} = \frac{1}{\sqrt{\Lambda}}.
\end{equation}
Hence, since the event horizon and the critical radius of an extreme SdS BH coincide, the local flow for such system starts immediately outside the BH. 

In addition, the information (or entropy) of an extreme SdS BH is $I_{BB} = \frac{5 \pi c^3}{\hbar G \Lambda ln2}$ \citep{GSbh}, which means that its formation via any astrophysical process is forbidden, as it would violate the second law of thermodynamics. Thus, by considering this  and Eqs.(\ref{hor}, \ref{SdSm}, \ref{horcrit}), we can conclude that the extreme SdS BH is the absolute upper theoretical limit for all possible systems with the maximum value for the local Hubble flow: 
\begin{equation}
H^2= \Lambda c^2.
\end{equation}

On the other hand, we can perform the same analysis for the theoretical lower limit. Specifically, by taking Eqs.(\ref{cosmic}, \ref{llimit}) into account, it becomes clear that the dS solution is the absolute lower limit for the local flow.

Thus, considering all the above points, we can state that the extreme SdS BH and dS are the theoretical upper and lower limits for the local Hubble flow. In the next section we  study these two limits in order to obtain the evolutionary process of the local Hubble flow.

\section{Free energy and maximum force--tension} 

In this section we obtain the free energy of the extreme SdS and dS as a criterion to show the direction of evolution during the local Hubble flow. The free energy of a system is defined as
\begin{equation}\label{F}
F = E -TS
,\end{equation}   
where $E$ is the total energy of the system, $T$ is the temperature, and $S$ is the entropy. For both cases, SdS and dS, the temperature according to  \citet{HG} is
\begin{equation}\label{T}
T = \frac{\hbar g}{2 \pi c k_B},
\end{equation}
where $g$ is the surface gravity, and for extreme SdS at $r_{crit}$ and dS at $\sqrt{\frac{3}{\Lambda}}$ is 
\begin{equation}\label{surSdS}
g_{SdS} |_{r= \frac{1}{\sqrt \Lambda}} = 0, \quad g_{dS} |_{r= \sqrt{\frac{3}{\Lambda}}} = \frac{c^2 \sqrt{\Lambda}}{\sqrt{3}}. 
\end{equation}
Taking all the above points into account we can obtain the free energy:  
\begin{equation}\label{Flimit}
F_{SdS}= \frac{c^4}{3 G \sqrt{\Lambda}}, \quad F_{dS} = 0. 
\end{equation} 
Table \ref{tab1} shows the corresponding quantities as well as the value of the Hubble constant for these two limits.

\begin{table}
\centering
\scalebox{0.8}{
\begin{tabular}{ |p{2.4cm}|p{2.1cm}|p{1.2cm}|p{1.8cm}|p{1.8cm}| }
\hline

Absolute limits & $H$ (km/s)/(Mpc)& Entropy &  Temperature & Free energy  \\ \hline
\hline
de Sitter&   $\sqrt{\frac{\Lambda c^2}{3}} = 56.1658$ & $\frac{3 \pi c^3 k_B}{G \Lambda \hbar}$ & $\frac{c \hbar \sqrt{\Lambda}}{2 \pi \sqrt 3 k_B}$ & $0$ \\ \hline
Extreme SdS&  $\sqrt{\Lambda c^2}=97.2821$ & $\frac{5 \pi c^3 k_B}{G \Lambda \hbar}$ & 0 & $\frac{c^4}{3 G \sqrt{\Lambda}}$ \\ \hline
\hline
\end{tabular}
}
\caption{The physical quantities for the theoretical upper and lower limits of the local H-flow}\label{tab1}
\end{table}

\noindent
Thus, the free energy  of any system where the local Hubble flow occurs is 
\begin{equation}\label{FEH}
\frac{c^4}{3 G \sqrt{\Lambda}} > F > 0,
\end{equation}
which, as can be easily checked, is in exact agreement with the notion of decrease in the amount of free energy in thermodynamics.

The maximum force--tension Barrow-Gibbons conjecture \citep{F1,F2} states that the maximum achievable force (or tension) in the Universe is equal to 
\begin{equation}\label{maxF}
F \leq \frac{c^4}{4G}.
\end{equation}
The argument is based on the fact that by considering the combination of three fundamental constants ($c$, $\hbar$, $G$) the obtained unit of the force (sometimes also called the Planck force) is free of $\hbar$: $F = \frac{c^4}{G}$. We also note that the notion of force is not defined in GR.

In the context of $\Lambda$-gravity the situation changes. Specifically, as stated above the presence of $\Lambda$ in the weak-field limit according to Eq.(\ref{force}) defines another limit for the weak- to strong-field transition. Consequently, in this new limit the self-gravitational force of the vacuum at $r= \sqrt{\frac{3}{\Lambda}}$ will be
\begin{equation}\label{Maxforce}
F_{max} = (\frac{\sqrt 3 c^2}{2 G \sqrt \Lambda}) (\frac{\Lambda c^2 r}{3})  |_{r=\sqrt {\frac{3}{\Lambda}}} = \frac{c^4}{2G}.
\end{equation}

The  obtained value is the maximum force and is two times larger  than the previously conjectured value in Eq.(\ref{maxF}). The important point is that this new limit is obtained based on the notion of {force} in the upper non-relativistic weak-field limit for de Sitter. 

At the same time, since the limit in Eq.(\ref{Maxforce}) is also the transition limit to relativistic de Sitter regimes, it is possible to show that the same value can be  obtained based on the relativistic notion of {tension,} which is defined as
\begin{equation}\label{Maxten}
F = \frac{mc}{t} = \frac{c^4}{2 G}.
\end{equation}

It should be noted that, generally in the Planck unit system $(c, \hbar, G),$ the notion of maximum force or Planck force is attributed to small scales and high energies, thus leading to the yet unknown quantum theory of gravity. However, regarding the linear nature of the  $\Lambda$-term in Eq.(\ref{force}), it turns out that the gravitational force (and energy) grows at very far distances and consequently another weak- to strong-field transition happens. Therefore, at the upper limit of non-relativistic regimes we obtain the notion of maximum {force}, which in contrast to other approaches does not have any conceptual problem. In addition, by transitioning to the relativistic regimes we can also obtain the $\frac{c^4}{2G}$, but this time according to standard definition of {tension}  established in  relativistic physics. 

For both cases it is clear that no quantum effect has a role. However, at the same time  we arrive at the highest possible achievable force--tension in non-relativistic and relativistic limits. So from the conceptual point of view,   the reason why there is no $\hbar$ in the definition of Planck force is that   it is reached in the non-quantum regimes, based on the first principles.

This notion of maximum tension is valid for all epochs of cosmological evolution \citep{GS4,GSbh}; for the characteristic mass and radius $(\frac{\pi c^8}{2 i \hbar G^2 H^3_i}, \frac{c}{H_i})$ the maximum value of tension remains constant and is equal to $\frac{c^4}{2G}$.

\section{Conclusions}

In this paper we studied the non-relativistic cosmology based on two different approaches by McCrea-Milne and by Zeldovich. The essential difference of these two approaches is  that  the former  is based on the {energy} notion, while the latter deals with the  notion of {force}. As a result, considering the two approaches plus the cosmological constant in weak-field GR, we obtained the full set of equations of non-relativistic cosmology, which are similar to FLRW equations in their form.

The two sets of equations,  Friedmann  and  non-relativistic  with $\Lambda$ as a constant of gravitational interaction, define two flows (global and local) with  different Hubble parameters. The latter can appear as the Hubble tension.

For the local Hubble flow we found the theoretical upper and lower limiting values for the Hubble parameter (i.e., for the extreme SdS and dS BHs). Then, we obtained the corresponding quantities for these limits and calculated the free energy for both of them. We concluded that for any system where the local flow has taken place the free energy must be between zero and  $\frac{c^4}{3 G \sqrt{\Lambda}}$. We analyzed the Barrow-Gibbons maximum force--tension conjecture and obtained a new limit that  is two times   larger than the previously conjectured value and that is conceptually self-consistent  with classical and relativistic notions of force and tension.


\end{document}